\documentclass[12pt]{iopart}

\usepackage{iopams}

\usepackage{graphicx}
\usepackage{epsfig}
\usepackage{dcolumn}
\usepackage{textcomp}
\usepackage{bm}

\newcommand{\field}[1]{\mathbb{#1}}

\newcommand{\ket}[1]{|#1\rangle}
\newcommand{\bra}[1]{\langle #1|}

\newcommand\T{\rule{0pt}{2.6ex}}
\newcommand\B{\rule[-1.2ex]{0pt}{0pt}}

\begin{document}

\title{Precision characterisation of two-qubit Hamiltonians via entanglement mapping} 

\author{Jared H. Cole\footnote[3]{Corresponding author's e-mail: j.cole@physics.unimelb.edu.au}, Simon J. Devitt and Lloyd C. L. Hollenberg}
\address{%
Centre for Quantum Computer Technology, School of Physics, The University of Melbourne, Melbourne, Victoria 3010, Australia.
}

\date{\today}
             
\begin{abstract}
We demonstrate a method to characterise the general Heisenberg Hamiltonian with non-uniform couplings by mapping the entanglement it generates as a function of time.  Identification of the Hamiltonian in this way is possible as the coefficients of each operator control the oscillation frequencies of the entanglement function.  The number of measurements required to achieve a given precision in the Hamiltonian parameters is determined and an efficient measurement strategy designed.  We derive the relationship between the number of measurements, the resulting precision and the ultimate discrete error probability generated by a systematic mis-characterisation.  This has important implications when implementing two-qubit gates for fault-tolerant quantum computation.
\end{abstract}

\maketitle

\section{Introduction}
One of the key requirements for a physical system to be used for quantum information processing applications is that the system must have a controllable two-qubit coupling\cite{Nielsen:00}\cite{Divincenzo:00}.  This is typically realised by an interaction between a pair of two-level systems which act as qubits.  It is this interaction which leads to entanglement and the `spooky action at a distance' effects which give quantum computers their power.  While some systems have a well defined  native two-qubit interaction, this is not generally the case.  In solid-state systems the interaction Hamiltonian is often a function of many control and fabrication parameters\cite{Koiller:02, Spoerl:05, Vrijen:00}.  As such, the form of the Hamiltonian can vary from device to device and even vary within different sections of a single device.  This means characterisation of some sort is critical in order to control the interaction and produce accurate gate operations for quantum computing applications.  

In this paper, we show how mapping the entanglement of the system as a function of time gives a conceptually straightforward approach to determining the dynamics of the system.  Specifically, we show how this method can be used to characterise a two-qubit interaction of the Heisenberg type,  
\begin{equation}\label{eq:H}
H=c_1XX+c_2YY+c_3ZZ,
\end{equation}
where $c_i\in\field{R}$, $XX=\sigma_x\otimes\sigma_x$ etc.\  and $\sigma_{i}$ are the Pauli operators.  Many solid-state quantum computing proposals rely on this type of interaction\cite{Kane:98, Loss:98, Vrijen:00, Friesen:03, Ardavan:03,Benjamin:03, Divincenzo:00b, Mohseni:05, Makhlin:01, Hollenberg:04}, as the general Heisenberg case covers a large class of quantum systems including real spin (i.e.\ exchange coupling) systems\cite{Kane:98, Loss:98, Friesen:03, Vrijen:00} and pseudo spin systems such as charge based designs\cite{Makhlin:01,Hollenberg:04}.  Recent work has also shown that two-qubit gates can be designed from a Heisenberg Hamiltonian with anisotropic couplings ($c_1\neq c_2\neq c_3$), as long as the components of the Hamiltonian are known accurately\cite{Wu:02}.  In an implementation of a quantum computer consisting of nominally identical qubits, the physical interaction between any given pair of qubits is similar, so we expect the structure of the Hamiltonian to be similar across a given device.  On the other hand, the size of the various couplings are a strong function of the fabrication process and will therefore vary from qubit to qubit.  In these situations, not only is it important to identify the size of the relative components, but for scalable systems this characterisation must be done in an \emph{efficient} manner, by which we mean that the process can be largely automated and require minimal physical modification to the original fabricated qubits. 

The issue of systematic, accurate and repeatable characterisation has far reaching consequences for quantum computing, given the ongoing efforts to define an error threshold, below which arbitrary quantum computation is possible using the concepts of concatenated quantum error correction and fault-tolerance\cite{Shor:96, Divincenzo:96b,Gottesman:98}.  Recent work has put this threshold at $10^{-2}-10^{-4}$ (depending on available resources) as the probability of a discrete gate error\cite{Alicki:06, Steane:03, Knill:05, Reichardt:04}, though this is the total error probability which is a combination of environmentally induced errors, characterisation and control errors.  By defining a systematic method of characterisation, we relate the number of measurements required in the initial characterisation phase to the resulting gate error rate, directly linking the required characterisation to the concatenated quantum error correction threshold.

Traditionally, characterisation has been performed using state and process tomography\cite{Poyatos:97, Chuang:97, James:01}, where a pulse sequence is developed to realise a certain gate, assuming the basic form of the Hamiltonian is known on experimental or theoretical grounds.  The effect of this gate on a complete set of input states is measured to build up the system state.  This  has been the method of choice for most early two-qubit experiments as the exact details of the interaction are not needed as long as the required two-qubit gate can be constructed approximately and the complete state of the system mapped.  This gives extensive information about the system including the effects of decoherence or loss channels.  If the gate is not ideal, then a good model is required, otherwise there is no systematic way of improving the performance of the gate or knowing whether an improvement is possible.  

A method for single-qubit characterisation has been recently developed which allows the efficient determination of the terms in the system Hamiltonian and can be implemented with minimal information about the system being characterised\cite{Schirmer:04, Cole:05, Cole:06}.  Rather than assuming knowledge about the system, this method involves mapping the system evolution over time and using this to gain information about the Hamiltonian itself.  While this typically requires many measurements to build up the evolution of the state of the system, it also provides detailed information about the form of the Hamiltonian.  This allows any necessary gate sequence to be developed offline without the need to tomographically map every gate that may be required in a given quantum circuit.

We show how the application of an accurately characterised Hadamard gate and measurement on both qubits is sufficient to find all the couplings in the Heisenberg Hamiltonian.  The result is that, using the machinery of a quantum computer architecture only, one can extract sufficient information to determine the fundamental interaction Hamiltonian and hence construct any required unitary gates.  By performing a combination of single- and two-qubit characterisation, the system can be `boot-strapped' from minimal knowledge of the system to provide all the required parameters for full controllability.

In contrast to spectroscopy, re-characterisation can be performed in-situ at any future time if required (e.g.\ to correct for long term drifts of the system parameters).  Addition characterisation steps can then be performed in parallel with the quantum computer's usual operation, whenever qubits are idle.

\section{Entanglement generated by the Heisenberg Hamiltonian}
Many two-qubit interactions can be described by the general Heisenberg Hamiltonian given in Eq.~(\ref{eq:H}).  When $c_1=c_2=c_3=d$, this is the conventional (isotropic) Heisenberg interaction of the form $H=d(XX+YY+ZZ)$, which is typical of spin based qubit coupling.  If $c_3=J$ and $c_1=c_2=0$, this is the interaction due to an Ising type coupling ($H=J ZZ$), common in pseudo spin schemes.  From the point of view of two-qubit gate design, for an Ising interaction it is not important which of the three terms is non-zero as the Hamiltonians $H=J XX$, $H=J YY$ and $H=J ZZ$ are locally equivalent\cite{Zhang:05}.  For this analysis we consider general Hamiltonians with $c_1$, $c_2$ and $c_3$ treated as parameters to be determined.  

We will restrict ourselves to only consider Hamiltonians which are piecewise constant in time and we assume controllability of any single qubit terms such that they can be turned off during the two-qubit interaction, or alternatively the single-qubit terms commute with the rest of the Hamiltonian.  The restrictions imposed by this assumption are discussed in section~\ref{sec:singlequbits}.  

As we are interested in reducing the systematic errors introduced by imperfect characterisation (rather than random errors caused by interaction with the environment), we have assumed that the effect of decoherence is negligible within the observation time\cite{Cole:06}.

To begin, we analytically derive the evolution of the system described by Eq.~(\ref{eq:H}) from some initial state $|\psi(0)\rangle$ to the state $|\psi(t)\rangle$ at some later time.  This evolution will, in general, depend on both the initial state and the components of the Hamiltonian.  To measure this evolution, the simplest method is to repeatedly initialise the system in $|\psi(0)\rangle$, allow the system to evolve for a time $n\Delta t$ and then measure the system.  This process is then repeated for integer values of $n$ to build up the time evolution at discrete time steps separated by $\Delta t$.  The difficulty with this process is that the time evolution is both a function of the single qubit terms and two qubit terms in the Hamiltonian.  

Alternatively, we can look at the entanglement generated by the interaction.  By definition, if the entanglement changes with time, then a two qubit interaction must be present, since local operations alone cannot generate a change in entanglement\footnote{While a change in entanglement can be used to infer the existence of two-qubit interaction terms, it cannot be used to exclude the presence of single qubit terms within the Hamiltonian.}.  This leads us to the idea of using the variation in the entanglement to analyse the interaction and isolate the effect of the terms of interest in the Hamiltonian.

The entanglement of the state generated by this evolution can be quantified using the \emph{squared concurrence}\cite{Wootters:98}
\begin{equation}
C^2=|\bra{\psi^*}YY\ket{\psi}|^2,
\end{equation}
where $C^2$ varies between $0$, when the qubits are unentangled, to $1$ when they are maximally entangled.  One method of measuring the concurrence is to measure the system in the $ZZ$ and $XZ$ bases.  We write the probability of measuring the $i\rm{th}$ qubit in the $\lambda_i$ eigenstate of the $\alpha_i$ operator as $
P^{\lambda_1\lambda_2}_{\alpha_1\alpha_2}$, where $\lambda_i =
\pm 1$ and $\alpha_i = X,Z$.  For example, in conventional notation this gives
\begin{equation}
P^{+-}_{ZZ}=|\langle01\ket{\psi}|^2
\end{equation}
or
\begin{equation}
P^{++}_{XZ}=\frac{1}{2}|\langle00\ket{\psi}+\langle10\ket{\psi}|^2.
\end{equation}
In terms of these quantities the squared concurrence is given by:\cite{He:03,Sancho:00}
\begin{equation}\label{eq:conc1}
C^2 = 4\left[P^{-+}_{ZZ}P^{+-}_{ZZ} + P^{--}_{ZZ}P^{++}_{ZZ} - 2
{\sqrt{\prod_{ij} P^{ij}_{ZZ}}}\cos(A+B)\right]
\end{equation}
where
\begin{equation}\label{eq:conc2}
\cos(A) = {{2 P^{++}_{XZ}- P^{--}_{ZZ}- P^{-+}_{ZZ}}\over {2 \sqrt
{P^{--}_{ZZ}P^{-+}_{ZZ}}}}
\end{equation}
and
\begin{equation}\label{eq:conc3}
\cos(B) = {{2 P^{-+}_{XZ}+ P^{--}_{ZZ}+ P^{-+}_{ZZ}-1}\over {2 \sqrt
{P^{+-}_{ZZ}P^{++}_{ZZ}}}}.
\end{equation}

In Table~\ref{tb:timeevolEnt} we consider the time evolution of the entanglement given four different initial states ($|\psi_1\rangle$ to $|\psi_4\rangle$).  In each case the evolution is a simple sinusoidal function with frequency given by the combination of two of the three parameters in the Hamiltonian given in Eq.~(\ref{eq:H}).
\begin{table}[tb!]
\begin{tabular}{| c | c | c |} \hline
Input	 State & $|\psi(0)\rangle$ \T \B	& $C^2(t)$ \\ \hline \hline
$|\psi_1\rangle$ & $|00\rangle$ \T \B	& $\sin^2[2(c_1-c_2)t]$  \\ \hline
$|\psi_2\rangle$ & $|01\rangle$ \T \B	& $\sin^2[2(c_1+c_2)t]$  \\ \hline
$|\psi_3\rangle$ & $(|0\rangle+ |1\rangle)\otimes(|0\rangle+ |1\rangle)$ \T \B	& $\sin^2[2(c_2-c_3)t]$	 \\ \hline
$|\psi_4\rangle$ & $(|0\rangle+ |1\rangle)\otimes(|0\rangle- |1\rangle)$ \T \B	& $\sin^2[2(c_2+c_3)t]$	 \\ \hline
\end{tabular}
\caption{The analytic form of the entanglement generated by Eq.~(\ref{eq:H}) for four different input states.\label{tb:timeevolEnt}}
\end{table}

Using the set of input states $|\psi_1\rangle$ to $|\psi_4\rangle$, the evolution of the system due to the Heisenberg Hamiltonian results in a significant simplification of Eqs.~(\ref{eq:conc1})-(\ref{eq:conc3}).  For instance, if the systems starts in state~$|\psi_1\rangle$, then $P^{+-}_{ZZ}=P^{-+}_{ZZ}=0$ for all time, whereas starting with~$|\psi_2\rangle$ gives $P^{++}_{ZZ}=P^{--}_{ZZ}=0$.  In fact, for~$|\psi_3\rangle$ and~$|\psi_4\rangle$, $P^{++}_{ZZ}=P^{+-}_{ZZ}=P^{-+}_{ZZ}=P^{--}_{ZZ}=1/4$ and therefore these states need not be measured at all.  These relations drastically reduce the number of measurements required to determine the concurrence and are true for any value of the coefficients of Eq.~(\ref{eq:H}), as they lead directly from the symmetries of this Hamiltonian.

The input states considered here are either the computational states or can be reached from the computational states using a Hadamard rotation on both qubits.  As the frequency of oscillation in each case is a linear combination of the coefficients $c_i$, determining the frequencies for evolution from the four starting states determines all the parameters including their signs.  The choice of which input states to use is largely arbitrary, depending on which frequency components are to be measured and which states can be prepared most easily.  The four states discussed here are chosen purely for the fact that they can be prepared from the computational states using only Hadamard gates.

A side effect of using the Fourier transform and the squared concurrence is that it removes any sign information, hence the need for four states in general.   If the sign of all the coefficients are known beforehand, or can be determined with a minimal number of measurements, then any three of these input states are sufficient for complete characterisation. 

The Fourier transform of the oscillation data gives the system parameters but, in contrast to the single-qubit case\cite{Cole:05}, these depend on the peak positions in frequency space, rather than the peak amplitudes.  While the oscillation frequencies present in the concurrence evolution are also present in the original probability evolution, the use of entanglement as a measure means the evolution is invariant under interchange of qubits and unaffected by the inclusion of single qubit terms which commute with the two-qubit interaction.

At this point an obvious question is, can we use other measures of entanglement or is concurrence somehow special?  As we are only considering pure states, all bipartite entanglement measures are equivalent and so the difference comes down to implementation.  In order to measure the Hamiltonian components accurately, it is important that the entanglement measure we use does not artificially introduce spurious frequencies into the evolution.  This immediately rules out any entropic measure which depends on a function of the form $f(x)=x\log(x)$ because if $x(t)$ varies sinusoidally, the logarithm of this function contains an infinite number of higher order harmonics.  These higher order harmonics complicate the frequency analysis and prevent unambiguous discrimination of the Hamiltonian components.  Most common entanglement measures are in some way related to the von Neumann entropy (i.e.\ $x(t)\Leftrightarrow\rho(t)$) and therefore suffer from this problem.  These include the entropy of entanglement~\cite{Bennett:96}, the entanglement of formation~\cite{Wootters:98} and logarithmic negativity~\cite{Plenio:05}.  Interestingly though, using the square of the negativity itself as an entanglement measure results in equivalent expressions to those obtained with the square of the concurrence.

Another consideration is how easily can the required measurements be performed experimentally, as most measures of entanglement require the complete reconstruction of the density matrix or at least a partial reconstruction.  The advantage of using concurrence is that it has a closed form which requires only two measurement channels, as shown in Eqs.~(\ref{eq:conc1})-(\ref{eq:conc3}).  In fact this is the minimum number of measurement channels required to characterise a Heisenberg type Hamiltonian with arbitrary coefficients.

\section{Uncertainty estimation and gate errors}
\begin{figure} [tb!]
\centering{\includegraphics[width=6cm]{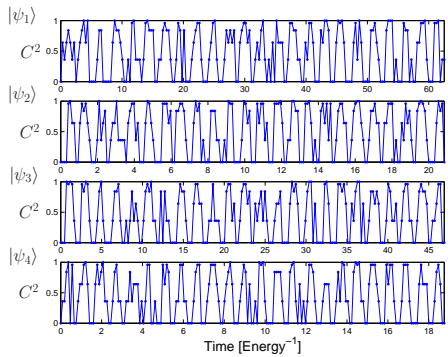}} 
\caption{Plot of the sampled entanglement as a function of time for the input states given in table~\ref{tb:timeevolEnt}, for an example Hamiltonian $H=1.2 XX + 0.6 YY + 1.4 ZZ$.  Each time point is the average of $N_e=10$ measurements and there are $N_t=200$ time points.  In each case the observation time has been chosen to obtain consistent sampling for each input state.\label{fig:timeexample}}
\end{figure}

\begin{figure} [tb!]
\centering{\includegraphics[width=6cm]{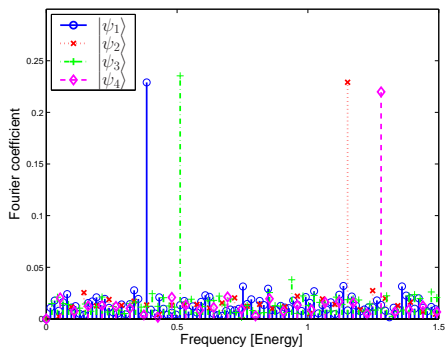}} 
\caption{Discrete Fourier transform of the data shown in Fig.~\ref{fig:timeexample} for different input states.  From the position of the peaks, the values of the Hamiltonian parameters can be determined.\label{fig:FFTexample}}
\end{figure}

To illustrate our analysis procedure visually, Fig.~\ref{fig:timeexample} shows the evolution of the entanglement for an example Hamiltonian given $N_e=10$ entanglement measurements at each time point.  Fig.~\ref{fig:FFTexample} shows the Fourier transform of this data, showing the peaks clearly above the noise floor.  From this example we see that even though the oscillations in the time domain are not well resolved, the peaks can clearly be seen above the discretisation (or `projection') noise in the frequency domain.

As this characterisation process ultimately relies on accurate determination of the oscillation frequency, many of the existing techniques for frequency standards are directly applicable\cite{Huelga:97, Wineland:94}.  Ultimately, there are two parameters to be chosen, the number of discrete time points, $N_t$, and the number of ensemble measurements, $N_e$.  The minimum number of discrete time points is governed by the Nyquist criteria, giving $N_t\ge2 t_{\rm{ob}}/t_{\rm{osc}}$ where $t_{\rm{osc}}$ is the period of oscillation and $t_{\rm{ob}}$ is the maximum time over which the system is observed.  To reduce the frequency uncertainty, $t_{\rm{ob}}$ should be maximised, though this will be limited by the decoherence time of the system.  As we have a single frequency oscillation, the uncertainty in the frequency determination can be reduced by having large numbers of ensemble measurements on the last few time points and using this to estimate the phase of the oscillation.  

In the ideal case (where $N_t$ is large), only two measurements are necessary at all time points with the exception that $N_e$ measurements are taken at the final two points, giving a total number of measurements $N=2N_t+2N_e$.  This is in contrast to the example given in Fig.~\ref{fig:timeexample}, where the same number of measurements are taken at each time point.  The error in the phase determination on the final two points is given by the projection noise and scales as $1/\sqrt{N_e}$, given the uncertainty in the frequency as $\delta f=2/(t_{\rm{ob}} \sqrt{N_e})$~\cite{Huelga:97}.  The fractional uncertainty in the frequency is then given by 
\begin{equation}\label{eq:dfbyf}
\frac{\delta f}{f}\ge\frac{4}{N_t\sqrt{N_e}}.
\end{equation}

While this analysis is quite straight forward, for quantum computing applications it is important to link these uncertainties to typical error models to determine the probability of a gate error produced by an uncertainty in the measured system Hamiltonian.  To do this we define an imperfect gate operation $U_{\rm{im}}=U_{\epsilon} U$ such that $U$ is the required gate operation followed by some error gate $U_{\epsilon}$.  Given $U_{\rm{im}}$, the effective error gate is $U_{\epsilon}=U_{\rm{im}}U^{-1}$.  The effective error probability is then defined as $p_{\rm{eff}}=1-|\rm{Tr}[U_{\epsilon}]/4|^2$.

If the Hamiltonian deviates from the form expected on theoretical grounds by such an amount the the error introduced by this deviation is larger than that due to characterisation uncertainties, we then use the measured Hamiltonian (rather than the theoretical one) to construct the gate.  For many Hamiltonians, a two-qubit gate can be constructed using, at most, three applications of the Hamiltonian together with single qubit rotations\cite{Zhang:03, Zhang:05}.  As our procedure measures the various terms in the Hamiltonian directly, it allows the construction of a pulse sequence to perform the required two-qubit gate, even when the Hamiltonian differs greatly from the theoretically expected form.  Using this type of gate construction, the error rate of the gate is now governed by the characterisation uncertainties alone.

To make this more concrete, we can calculate the $p_{\rm{eff}}$ for two common examples of native gates, assuming they are generated from an \emph{ideal} Hamiltonian (i.e.\ theoretical).  The analysis is similar for the case of a well characterised but \emph{non-ideal} Hamiltonian, though there is a cumulative effect if the two-qubit interaction is applied multiple times.  

For an ideal Ising Hamiltonian ($c_1=c_2=0$, $c_3=J$), the native gate is the CNOT gate, which can be constructed by applying the Ising Hamiltonian for a time $t_{\rm{gate}}=\pi/(4J)$ combined with appropriate single-qubit rotations.  Consider an example where characterisation is performed on the system, resulting in $c_1=c_2=0$ and $c_3=J\pm\epsilon$, with $\epsilon=\delta f/f$ the uncertainty in the peak position.  We then take an imperfect gate generated by a pulse of length $t=\pi(1+\epsilon)/(4J)$.  This gives $p_{\rm{eff}}=\sin^2(\pi\epsilon/4)$ as the effective error probability, assuming errors in the single qubit rotations are negligible.  Similarly, for an ideal isotropic Heisenberg Hamiltonian ($c_1=c_2=c_3=d$), the native entangling gate is the square-root-of-swap ($\sqrt{\rm{SWAP}}$).  Following the same procedure (assuming that the characterisation procedure leads to a common uncertainty $\epsilon$ in the peak positions) we obtain $p_{\rm{eff}}=3\sin^2(\pi\epsilon/4)/4$.

\begin{figure} [tb!]
\centering{\includegraphics[width=7cm]{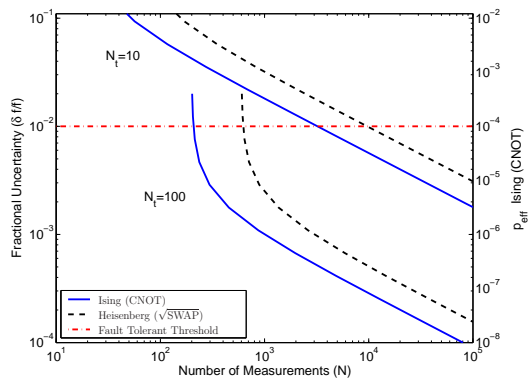}} 
\caption{The uncertainty in the Hamiltonian parameters as a function of the total number of measurements $N=2N_t+2N_e$, obtained from Eq.~\ref{eq:dfbyf}.  The curves are plotted for initial values of $N_t=10$ and $N_t=100$, for increasing $N_e$.  The right hand axis shows the effective probability of a discrete gate error ($p_{\rm{eff}}$) for the Ising case (the Heisenberg case differs by a factor of $3/4$).\label{fig:peff}}
\end{figure}

In Fig.~\ref{fig:peff}, $p_{\rm{eff}}$ is plotted for both the Ising and Heisenberg Hamiltonians for two different values of $N_t$ and compared to the conservative fault-tolerant threshold of $10^{-4}$.  The larger the value of $N_t$, the more precise the initial estimate when $N_e=1$.  As $N_e$ increases, the uncertainty scales as $1/\sqrt{N_e}$, as expected.  This allows us to calculate directly the time needed to initially characterise the system to obtain a given gate error rate.  For instance, if $N_t=10$ time points are chosen, then a conservative estimate of $N=10^4$ measurements are needed to reduce the error rate to below that required to satisfy the fault-tolerant threshold, again neglecting the effects of single qubit errors.  If more time points are used, the required number of measurements reduces accordingly, though this is limited by the requirement that at least two measurements are required at each time point to measure the concurrence.  These estimates for the number of measurements required should be compared to the case of single qubits\cite{Cole:05} where $N=10^4-10^8$ to achieve a probability of error, $p_{\rm{eff}}=10^{-4}$.

\section{Effect of single qubit terms}\label{sec:singlequbits}
Throughout this discussion, we have assumed that the unknown Hamiltonian took the form
\begin{equation}
H=H_{2q}+H_{1q},
\end{equation}
where $H_{2q}$ is given by Eq.~(\ref{eq:H}) and $H_{1q}$ are single qubit terms such that $[H_{2q},H_{1q}]=0$.  This restriction allows us to factor the evolution into separate single- and two-qubit evolution ($U=U_{2q}U_{1q}$) where the single qubit evolution $U_{1q}$ does not change the entanglement of the system.  While this may at first appear restrictive, it actually includes several Hamiltonians of interest to solid-state quantum computing.  This includes the effective exchange interaction between phosphorous donor spins in silicon\cite{Wellard:02} and the magnetic dipolar interaction between deep donors in silicon\cite{deSousa:04}.  For both these examples, the commutation relation holds, irrespective of the value of the various coupling parameters.  

A notable exception is the standard two-qubit interaction model for superconducting qubits\cite{Makhlin:01}.  In this case, not only is characterisation difficult but gate design is non-trivial and requires approximate and numerical methods\cite{Zhang:05}\cite{Spoerl:05}.  In general, for a Hamiltonian of arbitrary form, the eigenstates and therefore the evolution frequencies are non-linear functions of all the system parameters. 

In addition to single qubit terms which are part of the two-qubit interaction, we could also consider the effect of errors in the single qubit rotations used to prepare the input states given in Table~\ref{tb:timeevolEnt}.
In general the system evolution is a function of six frequencies given by the sum and differences of $c_1$, $c_2$ and $c_3$ and we have chosen the input states to isolated each frequency in turn.  Taking an imperfect input state which is close to one of the states given in Table~\ref{tb:timeevolEnt}, e.g.\ 
\begin{equation}
\ket{\psi_{1}(0)}\rightarrow\ket{\psi_{1}(0)}_{\rm{imp}}=\frac{1}{\sqrt{1+\eta}}(\ket{00}+\sqrt{\eta}\ket{01}),
\end{equation}
for some error probability $\eta$, and expressing the evolution of the concurrence in a series expansion about $\eta$, gives
\begin{eqnarray}
C^2(t) & = & \sin^2(2\omega_{1,-2} t)(1-2\eta) \nonumber \\
 & & +\frac{1}{2}\left[\cos(4\omega_{1,-3}t)-\cos(4\omega_{2,-3}t)\right. \\
 & & \left.+\cos(4\omega_{1,3}t)-\cos(4\omega_{2,3}t)\right]\eta+\mathcal{O}(\eta^2) \nonumber
\end{eqnarray}
where $\omega_{i,\pm j}=c_i\pm c_j$.  The evolution now contains oscillating terms at the other five system frequencies with amplitude $\eta$ as well as the original evolution at a frequency given by $c_1-c_2$ and amplitude $1-2\eta$.  As the Hamiltonian parameter estimates come from the position of the peak, the peak's position and therefore the estimate is unaffected by small errors in the input state.  

If the input state is completely unknown, the six frequency components are still present but there is now ambiguity as to which peak corresponds to which frequency.  The inclusion of imperfect alignment of the measurement bases has an similar effect to imperfect state preparation, with the amplitude of the undesirable frequency components now being related to the extent of the misalignment.  

We have not considered here the possibility of non-Heisenberg terms, such as $XZ$ or $YX$ as this complicates the situation considerably, again, introducing ambiguity into the frequency spectrum.  The effect of these terms is equivalent to a series of single qubit gates before and/or after the evolution\cite{Zhang:03, Zhang:05} and requires more sophisticated analysis\cite{Devitt:06}.  However, an upper bound on the size of these terms is again given by the projection noise and so scales as $1/\sqrt{N}$.

\section{Conclusion}
We have shown that mapping the entanglement generated by an unknown Hamiltonian provides a method of determining its structure and quantifying the various components.  The Heisenberg Hamiltonian has particularly nice properties which lead to an efficient method of characterisation by mapping the time evolution of the entanglement.  As this process requires finding the frequency of oscillation, the number of measurements required is typically much smaller than to precisely map the evolution of the expectation values.  The required input and measurement bases can be obtained using approximate Hadmard rotations only, which relaxes some of the requirements for accurate single qubit rotations as a precursor procedure.  In order to achieve precise control at, or below the fault-tolerant threshold the challenge is to be able to characterise logic gates to sufficient accuracy.  Given an uncertainty in the Hamiltonian parameters and using an effective error model, we determined the probability of error due to systematic mis-characterisation and this is linked directly to the error thresholds required for fault-tolerant quantum computation.  This type of characterisation procedure is of fundamental importance in experiments using two-qubit interactions, especially in the solid-state where precision control or uniformity of the Hamiltonian terms cannot be assumed \emph{a priori}.

\section*{Acknowledgments}
We would like to acknowledge helpful discussions with S.~G.~Schirmer, D.~K.~L.~Oi and A.~D.~Greentree.  This work was supported in part by the Australian Research Council, the US National Security Agency, the Advanced Research and Development Activity and the US Army Research Office under contract number W911NF-04-1-0290.  The authors thank the von Delft group at LMU for their hospitality and, for financial support, the DFG through the SFB631.  JHC and SJD acknowledge support from the Cambridge-MIT institute and LCLH was supported by the Alexander von Humboldt Foundation.

\section*{References}
\bibliographystyle{iopart-num}
\bibliography{hc2q}

\end{document}